\documentclass[12pt]{article}
\usepackage{epsfig,graphicx}
\usepackage{amsmath}

\title{Asymptotic freedom in strong magnetic field}
\author{M.A. Andreichikov, V.D. Orlovsky and Yu.A. Simonov,\\ Institute of Theoretical and Experimental
Physics\\ 117118, Moscow, B.Cheremushkinskaya 25, Russia}
\date{}

\newcommand{\be}{\begin{equation}}
\newcommand{\ee}{\end{equation}}

\def\fun#1#2{\lower3.6pt\vbox{\baselineskip0pt\lineskip.9pt
\ialign{$\mathsurround=0pt#1\hfil ##\hfil$\crcr#2\crcr\sim\crcr}}}

\newcommand{{\SD}}{\rm SD}

\newcommand{\vep}{\mbox{\boldmath${\rm p}$}}

\newcommand{\veQ}{\mbox{\boldmath${\rm Q}$}}

\newcommand{\veB}{\mbox{\boldmath${\rm B}$}}

\newcommand{{\Mc}}{\mathcal{M}}

\newcommand{\lan}{\langle}
\newcommand{\ran}{\rangle}

\begin{document}

\maketitle
\begin{abstract}
Perturbative gluon exchange interaction between quark and antiquark, or in a 3q
system, is enhanced in magnetic field and may cause vanishing of the total
$q\bar q$ or $3q$  mass, and even unlimited decrease of it -- recently called
the magnetic collapse of QCD. The analysis of the one-loop correction below
shows a considerable softening of this phenomenon due to $q\bar q$ loop
contribution, similarly to the Coulomb case of QED, leading to approximately logarithmic dumping of gluon exchange interaction ($\langle V \rangle \approx \mathcal{O}(1/\ln|eB|)$) at large magnetic field.
\end{abstract}

\section{}
Analysis of the hydrogen atom or positronium in strong magnetic field shows a
considerable enhancement of the Coulomb interaction, leading to the increase of
binding energy \cite{1,2,3,4}.This fact is due  to reduction of the system size
in the plane perpendicular to the direction of the magnetic field (MF) $\veB$,
making it closer
 to the one-dimensional  Coulomb system. As was shown  in \cite{5,2}, the binding energy in the leading
order in $\alpha$ grows as $ \ln^2 \left(\frac{B}{me^3}\right).$ It was shown
later, that the one-loop corrections to the one-photon exchange seriously
change the situation: in the hydrogen atom the binding energy tends to the
finite limit \cite{6,7}, while it shows  an unbounded growth in positronium
\cite{8}. One should note that the absolute value of binding energy in both
cases is not large and the upper limit of binding
 energy in hydrogen atom is 1.74~keV \cite{7,9} while  in positronium the collapse (vanishing)
of the total mass occurs at very strong fields: $B_{cr} \sim 10^{40}$ Gauss
\cite{8}.

Recently the dynamics of $q\bar q$ system in strong magnetic field was
studied in the framework of the relativistic Hamiltonian, derived from the path
integral for the corresponding  Green's function \cite{10}. The relevant
technic in the case of no MF was  extensively developed in \cite{11}. It was
shown, that the  one-gluon-exchange (OGE) interaction, or color Coulomb,
becomes increasingly important for large MF,  when OGE
 is taken in the leading (no quark loop) approximation. In particular, the mass of the $(q\bar q)$ meson
 vanishes at
$\sqrt{|e_q B|}\sim O(1$ GeV), i.e.for $B\approx 10^{19}-10^{20}$ Gauss.This
fact would imply a radical reconstruction of the vacuum, a proposal made in a
different context in \cite{12,13}.

A similar situation occurs in the case of baryons in strong MF: the baryon
(e.g. the neutron) mass
 vanishes at approximately the same $B_{\rm crit}$, as for mesons \cite{14}.

It is therefore very important to check whether the quark loop corrections may
stabilize the hadron mass at high MF, similarly to the case of the hydrogen
atom. As for gluon loop corrections, ensuring asymptotic freedom (AF), they are
neutral to MF, and AF only decreases the growth of binding energy (b.e.)
\cite{10} (b.e. grows as $\ln\ln \frac{eB}{\sigma}$ instead of $\ln^2eB$ in
atoms),  but does not prevent the collapse. But those are fermion loop
 contributions which stabilized hydrogen atom, and we shall below study  the quark-antiquark loops
in the case of the $q\bar q$ mesons, taking into account both confinement and
OGE interaction.

\section{} We start with the standard one-loop expression for the gluon self-energy part, which contributes to the gluon propagator as
\cite{15}\be D(q) = \frac{4\pi}{q^2-\frac{g^2(\mu^2_0)}{16\pi^2} \tilde \Pi
(q)}\label{1}\ee where $\tilde \Pi(q)$ contains the sum of gluon and  quark
loop terms, \be \tilde \Pi(q) = q^2 \Pi_{gl}(q)-\tilde \Pi_{q\bar q}
(q).\label{2}\ee

In absence of MF and neglecting strong interaction between gluons, one has \be
\Pi_{gl}(q) = - \frac{11}{3} N_c \ln \frac{|q^2|}{\mu_0^2},~~\tilde \Pi_{q\bar
q} (q) =-\frac23 n_f q^2 \ln \frac{|q^2|}{\mu^2_0},\label{3}\ee leading to the
standard AF expression for the OGE potential $(q^2=-\veQ^2 = -(q_\perp^2+q_3^2), ~~ \alpha_s ^{(0)}=
\frac{g^2(\mu^2_0)}{4\pi})$

\be V(Q) =-\frac43\frac{\alpha_s^{(0)}
4\pi}{Q^2\left(1+\frac{\alpha_s^{(0)}}{4\pi} \beta_0 \ln
\frac{Q^2}{\mu^2_0}\right)}=-\frac{16\pi}{3Q^2} \alpha_s(Q),~~ \alpha_s(Q) =
\frac{4 \pi}{\beta_0\ln \frac{Q^2}{\Lambda^2}},\label{4}\ee
where $\beta_0 = \frac{11}{3}N_c-\frac{2}{3}n_f$.

In the case of strong MF one can retain in $\tilde \Pi_{q\bar q}(q)$ the
contribution of  the lowest Landau levels (LLL), which couples only
to($q_0,0,0,q_3)$ polarizations and obtain the expression, known for a long
time \cite{16} for the $(e^+e^-)$ case, which is rewritten in our case by the
replacement $\alpha_{QED}\to \alpha_s ^{(0)}\frac{n_f}{2}$.

\be \frac{\alpha_s ^{(0)}}{4\pi}\tilde{\pi}_{q\bar q} (q) =- \frac{\alpha_s^{(0)}n_f|e_qB|}{\pi}\exp
\left(-\frac{q^2_\bot}{2|e_qB|}\right) T\left(
\frac{q^2_3}{4m^2}\right),\label{5}\ee where

$$T(z)=-\frac{\ln
\left( \sqrt{1+z}+\sqrt{z}\right)}{\sqrt{z(z+1)}} + 1
=\left\{\begin{array}{ll} \frac23z,&z\ll 1\\
1,& z\gg 1\end{array}.\right.$$

A convenient approximation with accuracy better than 10\% is  $T(z)
=\frac{2z}{3+2z}$\cite{7}.

\begin{figure}[t]
  \includegraphics[height=3.1cm]{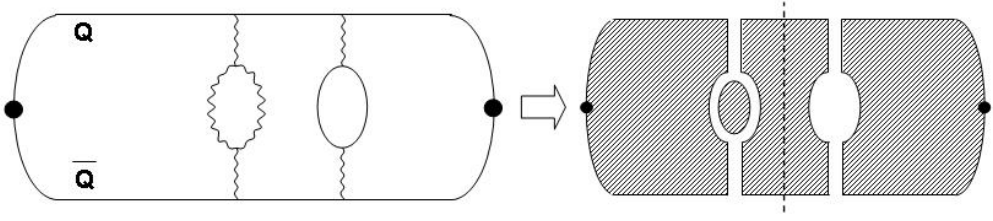}
      \caption{Gluon and $q\bar{q}$ loop insertions in the gluon exchange between quark $Q$ and antiquark $\bar{Q}$ in the meson $(Q\bar{Q})$. \label{fig1}}
\end{figure}

At this point one should  define the mass  parameter $m$, which in the case of
QED was (renormalized) electron mass \cite{6,7,8}. In our case the gluon and
quark loop contributions correspond to the graphs in Fig.1,
where we have denoted gluon line as a double quark line to make clear the gauge
interacting regions, and the confining regions are cross-hatched. One can see in Fig.1, that $q$  and
$\bar q$ in  the quark loop are not interacting  by simple gluon exchange
similarly to the $e^+e^-$ loop in the lowest order, but in the $q \bar q$ case
only the exchange of white objects (mesons or glueballs) can take place in
higher orders.

Moreover, quarks are moving on the borders of the confining surfaces and hence
should have the typical energies of quarks at the ends of the string -- they
are
 denoted as $\omega =\lan \sqrt{\vep^2_q + m^2_q}\ran$ in the path-integral Hamiltonian \cite{10,11}
and are of the order of $\sqrt{\sigma}, \sigma$ is string tension,
 $\sigma=0.18$ GeV$^2$. Thus one can replace $4m^2$ in (\ref{5}) by $4\sigma$.

Finally, one should take into account the nonperturbative (confining)
interaction inside the gluon loops, as shown in Fig 1. As shown in \cite{17}
this amounts
 to the replacement $\ln \frac {Q^2}{\mu^2_0} \to \ln\frac{Q^2+M^2_B}{\mu^2_0}$, where
$M_B\approx 1$ GeV and is expressed solely through $\sigma$. As  a result one
obtains the following form of the OGE interaction with account of gluon and
quark loop effects

\be V(Q) =-\frac{16\pi\alpha_s^{(0)}}
{3\left[Q^2\left(1+\frac{\alpha_s^{(0)}}{4\pi} \frac{11}{3} N_c \ln
\frac{Q^2+M^2_B}{\mu_0^2}\right)+ \frac{\alpha_s^{(0)}n_f |e_q
B|}{\pi}\exp\left(\frac{-q^2_\bot}{2|e_q B|}\right)T\left(\frac{q^2_3}{4\sigma}
\right)\right]}\label{6}\ee

where $\alpha_s^{(0)}=\frac{4\pi}{\frac{11}{3} N_c \ln
\frac{\mu_0^2+M_B^2}{\Lambda^2_V}}$, and $Q^2=q^2_\bot+q^2_3$.

\section{}

We  can now estimate the average value of $V(Q)$ in the meson state with the
wave  function, which takes into account magnetic field and confinement,
$V_{\rm conf} = \sigma  \eta$. The latter is convenient to replace by the
quadratic form $V_{\rm conf} \to \tilde V_{\rm conf} = \frac{\sigma}{2} \left(
\frac{\eta^2}{\gamma}+\gamma\right)$, with $\gamma$ to be found from the
stationary point condition, $\left.\frac{\partial M_{\rm mes}}{\partial
\gamma}\right|_{\gamma =\gamma_0}=0$. This replacement has accuracy of  the
order of 5\%, which is enough for our purposes. Then the $LLL$ wave functions
can be easily written \be \psi ( \eta_1,\eta_3) = \frac{1}{\sqrt{\pi^{3/2}
r^2_\bot r_3}}\exp \left( -
\frac{\eta^2_\bot}{2r^2_\bot}-\frac{\eta^2_3}{2r^2_3}\right),\label{7}\ee
where $r_\bot$ and $r_3$ are some functions of MF (see \cite{10} for details), for large fields
$r_\bot \approx \sqrt{\frac{2}{eB}}, \, r_3 \approx \sqrt{\frac{2}{\sigma}}$ and we can compute the OGE contribution to the meson mass $  \lan
V(Q)\ran_{mes}$, \be \lan V(Q)\ran_{mes} =\int V(Q) \psi^2 (q_\bot, q_3)
\frac{d^2q_\bot dq_3}{(2\pi)^3},\label{8}\ee
where $\psi^2 (q_\bot, q_3)$ is the Fourier transform of squared wave function $\psi^2( \eta_1,\eta_3)$.

 Insertion of (\ref{7}) and (\ref{6}) in (\ref{8})
yields
\be
\lan V(Q)\ran_{mes} =- C\int \frac{e^{-\frac{q^2_\bot r^2_\bot}{4} -
\frac{q^2_3r^2_3}{4}} d^2q_\bot dq_3}{Q^2A(q^2_\bot + q^2_3) + B(q^2_\bot ,
q^2_3)},\label{9}
\ee
where
 \begin{gather} A=1+ \frac{\alpha_s^{(0)}}{4\pi} \frac{11}{3} N_c \ln \left(
\frac{q^2_\bot + q^2_3 +M_B^2}{\mu_0^2}\right), \\ B=\frac{\alpha_s^{(0)} n_f
|e_qB|}{\pi}  e^{-\frac{q^2_\bot}{2|e_qB|}}T\left(\frac{q^2_3}{4\sigma}
\right), \quad C=\frac{16\pi\alpha_s^{(0)}}{3(2\pi)^3}. \label{10}
\end{gather}

\begin{figure}[h]
  \center{\includegraphics[height=7.0cm]{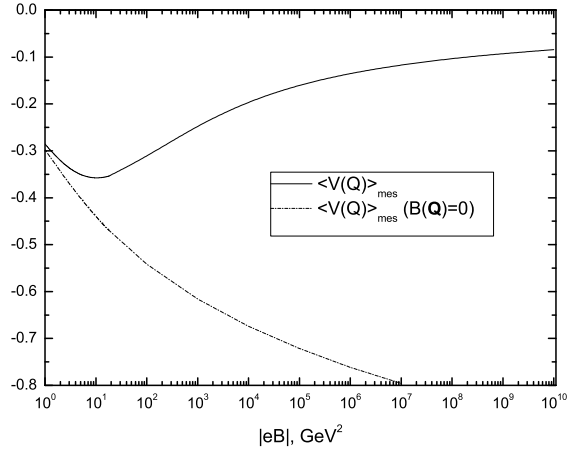}
    \caption{Coulomb correction to the meson mass (\ref{9}) in GeV as a function of magnetic field with (solid line) and without (broken line) account of quark loops contributions. \label{fig2}}}
\end{figure}

\begin{figure}[h]
  \center{\includegraphics[height=7.0cm]{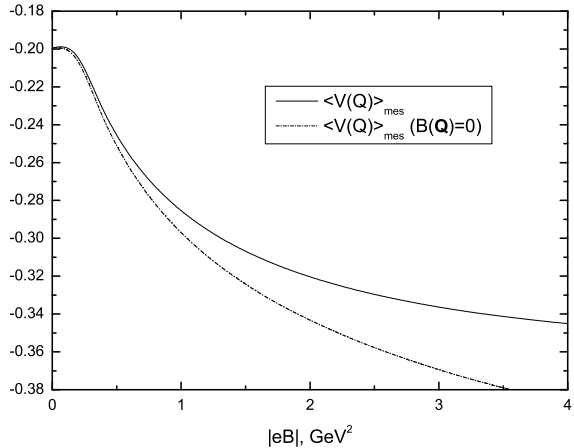}
    \caption{The same as in Fig.2, but for relatively small magnetic fields. \label{fig3}}}
\end{figure}

\section{}
Results of calculations for $\lan V(Q)\ran_{\rm mes}$ as a
function of MF are shown on Fig.2 for asymptotically large fields and on Fig.3 for relatively small fields. The values of parameters $\alpha_s^{(0)}$ and $\mu_0$ are connected by the relation
$\alpha_s^{(0)} = \frac{4\pi}{\frac{11}{3} N_c \ln \frac{\mu_0^2 +
M^2_B}{\Lambda_V^2}}$, and we have chosen $n_f=3$, $\mu_0 =1.1$ GeV, $\Lambda_V=0.385$ GeV, so $\alpha_s^{(0)} = 0.42$. As one can see from Fig.2 the account of quark loops contributions leads to the prevention of the so called magnetic collapse of QCD, -- the resulting correction vanishes at large MF (roughly as $-\frac{1}{\ln{|eB|}}$), so the meson mass is always finite.

\section{Acknowledgements}
The authors are grateful to B.O. Kerbikov for useful discussions. The financial support of Dynasty Foundation to V.D.O. is gratefully acknowledged.


\newpage



\begin{thebibliography}{99}
%

\bibitem{1}
L.I.Shiff and H.Snyder,  Phys. Rev. Lett.  {\bf 55}, 59 (1939).

\bibitem{2}
L.D.Landau and E.M.Lifshitz, Quantum Mechanism, Theoretical Physics, v.3,
Moscow, Fizmatlit, 2001, p.556.



\bibitem{3}
B.M.Karnakov and V.S.Popov, ZhETF {\bf 124}, 996 (2003); ibid  141, 5 (2012);



\bibitem{4}Jang-Haur Wang and Chen-Shiung Hsue, Phys. Rev. {\bf A 52}, 4508
(1995).




\bibitem{5} R.Loudon, Amer. J. Phys. {\bf 27}, 649 (1959); M.Andrews, Amer. J.
Phys. {\bf 34}, 1194 (1966).


\bibitem{6} A.E.Shabad and V.V.Usov, Phys. Rev. Lett. {\bf 98}, 180403 (2007);
Phys. Rev. {\bf D 77}, 125001 (2008).


\bibitem{7} M.I.Vysotsky, Pis'ma v Zh.Eksp. Teor. Fiz. {\bf 92}, 22 (2010);
B.Machet and M.I.Vysotsky, Phys. Rev. {\bf D 83}, 025022 (2011)

\bibitem{8} A.E.Shabad and V.V.Usov, Phys. Rev. Lett. {\bf 96}, 180401 (2006).






\bibitem{9}S.I.Godunov, B.Machet and M.I.Vysotsky, Phys. Rev. {\bf D 85}, 044058 (2012).

\bibitem{10}M.A.Andreichikov, B.O.Kerbikov and Yu.A.Simonov, arXiv:1210.0227 [hep-ph].

\bibitem{11}

 A.~M.~Badalian, B.~L.~G.~Bakker, and Yu.~A.~Simonov,  Phys.
Rev.  {\bf D 75}, 116001 (2007);
 A.Yu. Dubin,A.~B.~Kaidalov and   Yu.~A.~Simonov, Phys.~ Lett.~  {\bf B 323}, 41(1994);  Phys. At. Nucl. {\bf 56}, 1745 (1993); hep-ph/9311344.

\bibitem{12}J.Ambjorn and P.Olesen, Nucl. Phys. {\bf B315}, 606 (1989);  Phys.
Lett. {\bf B218}, 67 (1989).

\bibitem{13} M.Chernodub, Phys. Rev. Lett. {\bf 106}, 142003 (2011); Phys. Rev.
{\bf  D 82}, 085011 (2010).


\bibitem{14}~M.A.Andreichikov, V.D.Orlovsky and Yu.A.Simonov, (in
preparation).



\bibitem{15} M.E.Peskin and D.V.Schroeder, An Introduction to Quantum
Field Theory (Addison-Wesley, Reading, USA, 1995).


\bibitem{16}I.A.Batalin and A.E.Shabad, JETP, {\bf 33}, 483 (1971);
V.V.Skobelev, Russian Physics Jornal, v. 18, n. 10, 1481 (1975); Yu. M.Loskutov and
V.V.Skobelev, Phys. Lett. {\bf A 56}, 151 (1976); G.Calucci and R.Ragazzon, J.
Phys. {\bf A27}, 2161 (1994); A.V.Kuznetsov and  N.N.Mikheev, Phys. Rev. Lett.
{\bf 89}, 011601 (2002); hep-ph/0204201.

\bibitem{17}Yu.A.Simonov, Phys. At.Nucl. {\bf 74}, 1223 (2011); arXiv:1011.5386 [hep-ph]; Phys. At. Nucl. {\bf 58}, 107 (1995); hep-ph/9311247.




\end{thebibliography}
\end{document}